\documentclass[options,reqno]{amsart}
\usepackage{amsmath,amssymb}
\usepackage{geometry}
\usepackage{color}

\newcommand{\be}{\begin{equation}}

\newcommand{\ee}{\end{equation}}

\newcommand{\ba}{\begin{eqnarray}}
\newcommand{\ea}{\end{eqnarray}}
\newcommand{\baa}{\begin{eqnarray*}}
\newcommand{\eaa}{\end{eqnarray*}}
\newcommand{\bb}{\begin{thebibliography}}
\newcommand{\eb}{\end{thebibliography}}

\newcommand{\lab}[1]{\label{#1}}
\newcommand{\re}[1]{(\ref{#1})}

\renewcommand{\t}{\tilde}

\begin{document}
\title[Tridiagonal representations of the $q$-oscillator algebra]{Tridiagonal representations of the $q$-oscillator algebra and Askey-Wilson polynomials}

\author{Satoshi Tsujimoto}
\author{Luc Vinet}
\author{Alexei Zhedanov}

\address{Department of Applied Mathematics and Physics, Graduate School of Informatics, Kyoto University, Yoshida-Honmachi, Kyoto 606--8501, JAPAN}

\address{Centre de recherches math\'ematiques
Universite de Montr\'eal, P.O. Box 6128, Centre-ville Station,
Montr\'eal (Qu\'ebec), H3C 3J7, CANADA}

\address{
Department of Mathematics, School of Information, Renmin University of China, Beijing 100872, CHINA}

\begin{abstract}
A construction is given of the most general representations of the $q$-oscillator algebra where both generators are tridiagonal. It is shown to be connected to the Askey-Wilson polynomials.
\end{abstract}

\maketitle

\section{Introduction}
\setcounter{equation}{0}
The $q$-oscillator algebra \cite{Biedenharn,FV,JL2016,K,Mac} has two generators $A$ and $B$ that satisfy the relation
\begin{eqnarray}
\label{q-osc}
 AB - qBA =1.
\end{eqnarray}
It will be assumed that $q$ is not a root of unity. Relation \eqref{q-osc} provides a $q$-deformation of the commutation relation 
$[A,B]=1$ that is satisfied by the creation and annihilation operators of the quantum harmonic oscillator. This
$q$-algebra has found many applications and admit a standard representation in which $A$ and $B$ are separately upper and lower subdiagonal matrices.

We here construct the most general representations of the $q$-oscillator algebra in which $A$ and $B$ are non-degenerate tridiagonal operators. 
We also explain how these representations are related to the Askey-Wilson polynomials.

This study is in part suggested by analyses of ASEP probabilistic models which describe particles hopping on a line (segment) with the restriction, referred to as exclusion, that a particle cannot move to an occupied site.
Exact stationary solutions to such models with boundaries (where particles may enter and leave with certain probabilities)  can be obtained via the Matrix Ansatz approach \cite{Derrida1993}.
In this context, correlation functions are expressed in terms of matrices and vectors obeying certain constraints. 
As it turns out, for the matrices, these conditions are equivalent to the relation of the $q$-oscillator algebra.
(The vectors must be some generalized $q$-coherent states.)
In a seminal paper concerned with the ASEP model with general open boundaries, Uchiyama, Sasamoto and Wadati have exhibited pertinent realizations of the $q$-oscillator algebra where the generators are both tridiagonal. They have related these representations  to the Askey-Wilson polynomials and obtained exact results for quantities of interest in the model.
These observations have led to a beautiful combinatorial interpretation of the Askey-Wilson moments in terms of staircase tableaux \cite{CorteelWilliams2006, CSSW2012}

The aim of the present paper is to offer a constructive approach to these tridiagonal representations of the $q$-oscillator algebra and to explain, using an approach called tridiagonalization, the occurrence of the Askey-Wilson polynomials in connection with this simple $q$-algebra.

Let us note that subjecting two generators $A$ and $B$ to quadratic relations of the general form
\begin{eqnarray}
 \alpha_1 A^2 + \alpha_2 B^2 + \alpha_3 AB + \alpha_4 BA + \alpha_5 A + \alpha_6 B + \alpha_7 \mathbb{I}=0, \lab{quadrel}
\end{eqnarray}
has also found applications in kinetic  models \cite{ER} and martingale orthogonal polynomials \cite{BMW} where representations of the algebra \eqref{quadrel} have been explored.
It should be observed that in the generic situation, such quadratic algebras can be reduced \cite{OS,BMW} to the $q$-oscillator algebra under affine transformations of the type
$A \to \mu_1 A + \mu_2 B + \mu_0 \mathbb{I}$, $B  \to \nu_1 A + \nu_2 B + \nu_0 \mathbb{I}$.
We may finally remark that the tridiagonal representations of the Weyl relation $AB=qBA$ have been constructed in \cite{Z3}.  

The remainder of the paper proceeds as follows. In the next section, with the proviso that they be irreducible, the most general tridiagonal  $A$ and $B$ that verify  \eqref{q-osc} are constructed. These $A$ and $B$ are then identified with the Jacobi matrices associated to the big $q$-Jacobi polynomials. The tridiagonalization procedure is applied in Section 3 to the latter polynomials and employed to connect the Askey-Wilson polynomials to the $q$-oscillator algebra. 
In section 4, we consider two possible finite-dimensional reductions of the tridiagonal representation and the associated polynomials. 
The first reduction is well known and corresponds to the $q$-Hahn polynomials. The second reduction is new and leads to functions that we shall call the $q$-para Krawtchouk polynomials. The theory of the finite-dimensional representations of the $q$-oscillator algebra is brought to bear on these two cases. 
It is shown in particular that the $q$-para Krawtchouk polynomials correspond to  a reducible situation, {\it i.e.} that there exists a similarity transformation such that both matrices $A$ and $B$ can be presented as a direct sum of two irreducible matrices. 
In Section 5,  we consider the algebras behind the bispectrality of the polynomials that have arisen and relate them to the $q$-oscillator algebra.
First, we show that the operators $A$, $B$ and a diagonal operator $Z$ satisfy the commutation relations of a degenerate Askey-Wilson algebra, that can be called the big $q$-Jacobi algebra. Second we indicate that the operators $Z$, $L=A+\mu B$ and a third operator $M$ form the general algebra associated to the Askey-Wilson polynomials. 
The paper ends with some concluding remarks.

\section{Tridiagonal representations and big $q$-Jacobi polynomials}
\setcounter{equation}{0}
Let $e_n$, $n=0,1,\ldots$ be a fixed basis and assume that the two operators $A$ and $B$ are both tridiagonal with respect to this basis:
\be
A e_n = a_{n+1}e_{n+1} + b_n e_n + c_n e_{n-1}, \quad   B e_n = a_{n+1}' e_{n+1} + b_n' e_n + c_n'  e_{n-1} \lab{3_AB} \ee
with some coefficients $a_n, b_n, \dots, c_n'$. 
We assume that $A$ and $B$ are irreducible, namely that all the off-diagonal coefficients are non-zero:
\be
a_n c_n a_n' c_n' \ne 0, \quad n=1,2, \dots
\lab{irr} \ee
As a matter of fact, 
when one or more of the off-diagonal coefficients become zero, the matrix $A$ or $B$ can be
expressed as a direct sum of two or more independent Jacobi matrices.

The basis $\{e_n\}$ can be trivially changed as per $e_n \to \kappa_n e_n$ with arbitrary nonzero parameters $\kappa_n$. 
This leads to obvious transformations of the matrix coefficients $a_n , \dots, c_n'$.
Using this degree of freedom, we can choose parameters $\kappa_n$ such that $a_n =1$ for all $n=1,2,\dots$.  
We can thus take 
\be
A e_n = e_{n+1} + b_n e_n + u_n e_{n-1}, \quad   B e_n = \xi_{n+1}e_{n+1} + \eta_n e_n + \zeta_n u_n e_{n-1} \lab{AB_3diag1} \ee
without loss of generality.  
In the finite-dimensional case, the matrices $A$ and $B$ will be of the form
$$
A=\left[
\begin{array}{ccccc}
b_0 & u_1 & 0 & \cdots & 0 \\
1 & b_1 & u_2 &  & \vdots  \\
0 & \ddots & \ddots & \ddots & 0 \\
\vdots & & 1 & b_{N-1} & u_{N} \\
0 & \cdots & 0 & 1 & b_N
\end{array}
\right],\quad
B=\left[
\begin{array}{ccccc}
\eta_0 & \zeta_1 u_1 & 0 & \cdots & 0 \\
\xi_1 & \eta_1 & \zeta_2 u_2 &  & \vdots  \\
0 & \ddots & \ddots & \ddots & 0 \\
\vdots & & \xi_{N-1} & \eta_{N-1} & \zeta_N u_{N} \\
0 & \cdots & 0 & \xi_N & \eta_N
\end{array}
\right]. 
$$
We have then the following sets of unknowns:
$$
\{b_0, b_1, \dots, b_N\}, \; \{u_1, u_2, \dots, u_N\}, \: \{\eta_0, \eta_1, \dots, \eta_N\}, \; \{\xi_1, \dots, \xi_N\}, \; \{\zeta_1, \dots, \zeta_N\},
$$
and the irreducibility condition becomes 
\be
u_i \ne 0, \; \xi_i \ne 0, \; \zeta_i \ne 0, \; i=1,2,\dots, N. \lab{irr_u} \ee
In the infinite-dimensional case, all these sets are unrestricted from above.

Substituting \re{AB_3diag1} into \re{q-osc} and collecting the terms in front of $e_n$, we obtain the relation
\be
\Xi_{n+2}^{(1)} e_{n+2} +\Xi_{n+1}^{(2)} e_{n+1} + \Xi_{n}^{(3)} e_{n} + \Xi_{n}^{(4)} u_n  e_{n-1} + \Xi_{n}^{(5)} u_n u_{n-1} e_{n-2} =0 \lab{5_diag_0}, \ee
where 
\be
\Xi_{n}^{(1)}=\xi_{n-1}-q  \xi_{n}, \quad \Xi_{n}^{(5)}=\zeta_n  - q \zeta_{n-1},  \lab{xi_15} \ee
\be
\Xi_{n}^{(2)}= \xi_n (b_n-qb_{n-1}) +\eta_{n-1}-q\eta_n, \quad  \Xi_{n}^{(4)}=  \zeta_n (b_{n-1}-qb_{n}) + \eta_{n}-q\eta_{n-1},  \lab{xi_24} \ee
and 
\be
\Xi_{n}^{(3)} = \zeta_n u_n -q \zeta_{n+1} u_{n+1} + \xi_{n+1} u_{n+1} - q \xi_n u_n +(1-q) b_n \eta_n -1. \lab{xi_3} \ee
Relation \re{5_diag_0} should be valid for all $n=0,1,2,\dots$ and hence, for each $n$, we have 5 conditions
\be
\Xi_{n}^{(1,2)} =0, \; n=2,3,\dots, \; \Xi_{n}^{(2,4)} =0, \; n=1,2 \dots, \; \Xi_n^{(3)}=0, \: n=0,1,2,\dots \lab{xi=0} \ee
for 5 unknown coefficients $b_n, u_n, \xi_n, \eta_n,\zeta_n$.

Note that condition $\Xi_n^{(3)}=0$ for $n=0$ is
\be
u_1(\xi_1-q\zeta_1) +(1-q) b_0 \eta_0 =1, \lab{ini_Xi3} \ee
which formally can be written as $\Xi_0^{(3)}=0$ if one puts 
\be
u_0=0 \lab{u_0=0}. \ee

We will solve equations \re{xi=0} step-by-step starting with the elementary equations \re{xi_15}. 
From these equations we find
\be
\xi_n = \xi_0 q^{-n}, \quad \zeta_n = \zeta_0  q^{n}, \quad n=1,2,\dots \lab{xi_zeta} \ee
where $\xi_0, \zeta_0$ are  arbitrary nonzero constants.

Subtracting equations $\Xi_{n}^{(2)}=0$ and $\Xi_{n}^{(4)}=0$ from one another we have
\be
(1+q)(\eta_{n-1}-\eta_n) +z_n b_n - z_{n-1} b_{n-1}=0, \quad n=1,2,\dots, \lab{subs_be1} \ee
where
\be
z_n = \xi_n +q \zeta_n = \xi_0 q^{-n} + \zeta_0 q^{n+1}. \lab{z_n} \ee
From \re{subs_be1} we obtain the relation
\be
(q+1) \eta_n = z_n b_n -q s_1, \quad n=1,2,3,\dots  \lab{eta_b} \ee   
where 
\be
s_1 = (q^{-1} \xi_0 +  \zeta_0) b_0 -(1+q^{-1})\eta_0.
\lab{s_1} \ee
Similarly, also from the pair of equations \re{xi_24}, we have
\be
\xi_0 \zeta_0(1+q)(b_n-b_{n-1}) =z_n \eta_n - z_{n-1} \eta_{n-1}, \quad n=1,2,\dots \lab{subs_be2} \ee 
which leads to the relation
\be
(q+1) \xi_0 \zeta_0 b_n = z_n \eta_n -q s_2,
 \lab{b_eta} \ee 
with
\be
s_2= (q^{-1} \xi_0 +  \zeta_0) \eta_0 - \xi_0 \zeta_0 (1+q^{-1}) b_0. \lab{s_2} \ee
From equations \re{eta_b} and \re{b_eta}, we obtain explicit expressions for the coefficients $b_n, \eta_n$:
\be
b_n= \frac{s_2 (q+1) + s_1 z_n}{\gamma_n \gamma_{n+1}}, \quad n=1,2, \dots \lab{b_n} \ee
and
\be
\eta_n= \frac{s_1 \xi_0 \zeta_0(q+1) + s_2 z_n}{\gamma_n \gamma_{n+1}}, \quad n=1,2,\dots \lab{eta_n} \ee
where
\be
\gamma_n = \xi_0 q^{-n} - \zeta_0 q^n = (1-q)^{-1} (z_n - z_{n-1}). \lab{gamma_m} \ee
It should be stressed that the initial values $b_0, \eta_0$ (or, equivalently, the constants $s_1,s_2$) can be chosen arbitrarily.

We thus have found the explicit expressions of the diagonal coefficients $b_n$ and $\eta_n$. 
There only remains to find the expression for the off-diagonal matrix elements $u_n$. 
This can be done from the remaining condition \re{xi_3} which leads to the equation
\be
y_{n+1} u_{n+1} - y_{n-1} u_n = 1-(1-q)b_n \eta_n, \lab{eq_u} \ee 
where
\be
y_n = \xi_0 q^{-n} - \zeta_0 q^{n+1} = q^{1/2} \gamma_{n+1/2}. \lab{y_gamma} \ee
Multiply both sides of \re{y_gamma} by the factor $y_n$ to find
\be
V_{n+1} - V_n = y_{n} + (1-q) y_n b_n \eta_n, \lab{V_eq} \ee  
where
\be
V_n = y_n y_{n-1} u_n. \lab{V_u} \ee
The terms on the right hand side of \re{V_eq} can be presented in the same form as those on the left hand side of the same equation. 
Indeed, we have
\be
y_n = \frac{z_{n+1/2}-z_{n-1/2}}{q^{-1/2} - q^{1/2}} \lab{y_zz} \ee
and
\be
(q-1) y_n b_n \eta_n = K_{n+1} - K_n, \lab{yb_K} \ee  
where
\be
K_n = \frac{q^{2-n}(s_2+s_1\zeta_0 q^n)(s_2q^n + s_1\xi_0)}{\gamma_n^2}. \lab{K_n} \ee
The general solution of \re{V_eq} can thus be presented in the form 
\be
y_n y_{n-1} u_n = \frac{\xi_0 q^{-n} +\zeta_0 q^n}{q^{-1} - 1} + K_n + s_0, \lab{u_n_sol} \ee
where $s_0$ is an arbitrary constant.

If one demands that $u_0=0$ (this is equivalent to  the condition that the Jacobi matrix $A$ is semi-infinite), the constant $s_0$ is then fixed by this condition and one finds:
\be
s_0 = \frac{q \: \left(   (\xi_0+\zeta_0)(\xi_0-\zeta_0)^2  + q(1-q) \left(\xi_0 \zeta_0 s_1^2 +(\xi_0+\zeta_0)s_1 s_2 + s_2^2  \right) \right)}{(q-1)(\xi_0-\zeta_0)^2}. \lab{s_0} \ee
We thus obtain for $u_n$ the following expression
\be
u_n = \frac{1}{(1-q)\xi_0} \: \frac{q^n (1-q^n) (1-\tau q^n) \prod_{i=1}^4(1-\beta_i q^n) }{(1-\tau q^{2n})^2 (1-\tau q^{2n-1})(1-\tau q^{2n+1})}, \lab{u_fin} \ee  
with
\be
\tau=\zeta_0/\xi_0 \lab{tau_rep} \ee
and where the coefficients $\beta_i$ depend on the representation parameters $\zeta_0, \xi_0, s_1,s_2$ and must moreover satisfy the conditions:
\be
\beta_1 \beta_2 = \beta_3 \beta_4 = \tau =\zeta_0/\xi_0. \lab{res_beta} \ee
The main observation is that one can identify the matrix $A$ with the Jacobi matrix associated to the big $q$-Jacobi polynomials \cite{KLS}.
Indeed, let us introduce the operator $J$ defined by
\be
J e_n =  u_{n+1}^{(J)} e_{n+1} + b_n^{(J)} e_n + e_{n-1}, \quad Je_0 = u_1 e_1 + b_0 e_0, \lab{A_J} \ee
where
\be 
u_n^{(J)} = D_{n-1}C_n, \quad b_n^{(J)} = 1-D_n-C_n \lab{ub_J} \ee
and 
\begin{eqnarray}
 &&D_n = \frac{(1-c_1 q^{n+1})(1-c_1 c_2 q^{n+1})(1-c_3 q^{n+1})}{(1-c_1 c_2 q^{2n+1})(1-c_1 c_2 q^{2n+2})}, \nonumber\\
 &&C_n = -c_1c_3 q^{n+1}\frac{(1-q^n)(1-c_2 q^{n})(1-c_1 c_2c_3^{-1} q^{n})}{(1-c_1 c_2 q^{2n+1})(1-c_1 c_2 q^{2n})}, \lab{DC_J}
\end{eqnarray}
with $c_1,c_2,c_3$ arbitrary parameters.

This operator $J$ entails the recurrence relation of the big $q$-Jacobi polynomials. More precisely, let us introduce the column vector 
\be
\vec P = (P_0(x;c_1,c_2,c_3), P_1(x;c_1,c_2,c_3), \dots, P_n(x;c_1,c_2,c_3), \dots )^T, \lab{P_vec} \ee 
where $P_0(x) =1, P_1(x) = x-b_0^{(J)}$ and where $P_n(x;c_1,c_2,c_3)$ are the monic big $q$-Jacobi polynomials \cite{KLS} defined by the recurrence relation
\be
P_{n+1}(x) + b_n^{(J)} P_n(x) + u_n^{(J)} P_{n-1}(x) = x P_n(x). \lab{big_Jrec} \ee
Explicitly, these polynomials have the following explicit expression \cite{KLS}
\be P_n(x;c_1,c_2,c_3) = \mu_n \: {_3}F_2\left({ q^{-n}, c_1c_2q^{n+1},x \atop c_1 q, c_3 q};q \right); \lab{P_big_hyp} \ee
the normalization factor $\mu_n$ is chosen so that $P_n(x) = x^n + O(x^{n-1})$.  
We shall not need the precise formula for $\mu_n$ in the following.

The vector $\vec P$ is the eigenvector of the operator $J=A^T$:
\be
J {\vec P} = x {\vec P}. \lab{A_} \ee 
where $A^T$ means the transposed $A$.

Comparing the representation matrix element $u_n,b_n$ with the recurrence coefficients $u_n^{(J)}, b_n^{(J)}$, we conclude that they coincide up to a scaling transformation:
\be
u_n = \sigma^2 u_n^{(J)} , \quad b_n = \sigma b_n^{(J)} \lab{ub_kap} \ee
with a parameter $\sigma$ that can be found from the relation
\be
 (q-1) \xi_0 c_1 c_3 q \sigma^2 =1. \lab{sig_xi} \ee
Note that the operator $B$ is similarly related to the big $q$-Jacobi polynomials. 
Indeed, one can always perform a similarity transformation $B \to S^{-1} B S$ with $S$ a diagonal matrix, so that in the basis $\{e_n\}$, the operator $\tilde B$ acts as follows:
\be
\tilde B e_n  = S^{-1} B S e_n = e_{n+1} + b_n^{(B)} e_n + \tilde u_{n+1}^{(B)} e_{n+1}. \lab{B_e_til} \ee
The matrix elements $b_n^{(B)}$ and $u_n^{(B)}$ will also coincide with the recurrence coefficients of the big $q$-Jacobi polynomials up to the scaling
\be
 b_n^{(B)} = \kappa \tilde b_n  \quad u_n^{(B)}= \kappa^2 \tilde u_n, \lab{bu_B} \ee
where $\kappa c_3 q (q-1)=1$; here $\t b_n, \: \t u_n$ denote the recurrence coefficients of the big $q$-Jacobi polynomials with $\t c_1 = c_2, \t c_2 = c_1, \t c_3 = c_3$.
This means that the eigenvectors of the operator $B$ are the big $q$-Jacobi polynomials that are obtained from $P_n(x;c_1,c_2,c_3)$ by  interchanging the parameters $c_1 $ and $c_2$ and scaling the variable; in other words,
$B \vec Q = x \vec Q$ 
where $\vec Q =(Q_0, Q_1, \dots, Q_n, \dots)^{T}$ and 
$Q_n = \tilde \mu_n P_n(\kappa^{-1} x; c_2, c_1, c_3)$.

We thus have found that both operators $A$ and $B$ correspond to (the transposed of) the Jacobi matrix of the big $q$-Jacobi polynomials.

\section{From big $q$-Jacobi polynomials to Askey-Wilson polynomials}
\setcounter{equation}{0}
In the light of the previous section, we shall from now on assume that the operator $A$ coincides with the standard big $q$-Jacobi tridiagonal operator \re{A_J}. This is convenient for further analysis. 
Note that the $q$-oscillator algebra relations are unchanged under the substitution $(A,B) \to (B^T,A^T)$. 
We shall thus work with the parameters $c_1,c_2,c_3$ instead of the parameters $\xi_0, \zeta_0, s_1, s_2$ used so far 
in the general tridiagonal solution of the $q$-oscillator relation.

In this section we will show how the generic Askey-Wilson Jacobi operator can be obtained from the big $q$-Jacobi operator $A$ via the tridiagonalization procedure \cite{IK1,IK2,GIVZ}.

Recall that the big $q$-Jacobi polynomials satisfy the following $q$-difference equation
\be
E(x) P_n(xq) + F(x) P_n(x/q) -(E(x) + F(x) -c_1c_2q-1)P_n(x) = z_n P_n(x), \lab{dfr_J} \ee
where
\be
z_n =c_1 c_2 q^{n+1} + q^{-n} \lab{z_J} \ee
and
\be
E(x) = \frac{c_1 q (x-1)(c_2x-c_3)}{x^2},\quad F(x) =\frac{(x-c_1q)(x-c_3q)}{x^2}. \lab{BE_J} \ee
This motivates the introduction of an auxiliary diagonal operator $Z$ to supplement to the operator $A$, that will be taken to act
as follows in the basis $\{e_n\}$:
\be
Z e_n = z_n e_n , \quad n=0,1,2,\dots. \lab{Z_J} \ee
The method of tridiagonalization consists in  using $A$ and $Z$ to form another operator, again denoted by $B$, that will read
\be
B= r_1 ZA - q r_1 AZ + r_0 \mathbb{I}, \lab{B_A_J} \ee
where $\mathbb{I}$ is the identity operator and where
\be
r_0 =\frac{c_1(c_2+1)+c_3(c_1+1)}{c_1 c_3 (1-q^2)}, \quad r_1= -\frac{1}{c_1 c_3 q (q+1)(1-q)^2}.  \lab{r_01_J} \ee
By construction, $B$ is tridiagonal in the basis $\{e_n\}$.
This realization is in fact equivalent to the most general nondegenerate tridiagonal representation of the $q$-oscillator algebra constructed in the previous section.

The representation  in the ``recurrence picture'' with  $A=J$ (see \re{A_J}), $Z$ defined by \eqref{Z_J} and $B$ given by \re{B_A_J} is convenient because it is directly connected with the big $q$-Jacobi polynomials.
In the dual ``difference equation picture'', according to \eqref{dfr_J}, $Z$ is  given by the $q$-difference operator
\be
Z = E(x) T^+ + F(x) T^- -(E(X) + F(x)-c_1c_2q-1)\mathbb{I}, \lab{Z_dual} \ee
where $T^+ f(x) = f(xq), \: T^- f(x) = f(x/q)$
and $A$ is multiplication by $x$:
\be
A = x. \lab{A_x} \ee
The operator $B$ becomes $q$-difference operator
\be
B = \frac{(x-qc_1)(x-qc_3)}{q^2(q-1)c_1c_3x} T^- + \frac{1}{(1-q)x} \mathbb{I}. \lab{B_dfr} \ee
This is simply related to the well-known realization of the $q$-oscillator algebra in terms of the $q$-derivative operators,
where $A=x$ and $B=x\,T^{-} + 1/(1-q)$.

We now relate in this framework, the big $q$-Jacobi polynomials to those of Askey and  Wilson.
Recall that the Askey-Wilson polynomials depend on 4 parameters $a_1,a_2,a_3,a_4$ and satisfy the recurrence relation \re{big_Jrec}, where
\be
u_n^{(AW)} = \frac{1}{4} D_{n-1} C_n, \quad b_n^{(AW)} =  \frac{1}{2} \left(a_1 + a_1^{-1} - D_n - C_n \right), \lab{ub_AW} \ee 
with
\ba
&D_n = \dfrac{(1-a_1a_2 q^n)(1-a_1a_3 q^n)(1-a_1a_4 q^n)(1-g q^{n-1})}{a_1(1-g q^{2n-1})(1-g q^{2n})}, \nonumber \\
&C_n = \dfrac{a_1(1-q^n)(1-a_2a_3 q^{n-1})(1-a_2a_4 q^{n-1})(1-a_3a_4 q^{n-1})}{(1-g q^{2n-1})(1-g q^{2n-2})},
\lab{DA_AW}
\ea
and $g=a_1a_2a_3a_4$; the initial conditions are the standard ones $P_0=1, P_1(x) = x-b_0$.

Let us consider the  following operator
\be
W= \tau_1 ZA + \tau_2 AZ + \tau_3 A + \tau_0 \mathbb{I}, \lab{W_A} \ee
where $A$ and $Z$ are as above and $\tau_1, \tau_2, \tau_3$ are arbitrary parameters.
$W$ acts tridiagonaly in the basis $\{e_n\}$:
\ba
&&W e_n = (\tau_1 z_{n-1} + \tau_2 z_n + \tau_3) e_{n-1} +   (\tau_1 z_{n+1} + \tau_2 z_n + \tau_3) u_{n+1} e_{n+1} + \nonumber \\
&& \left( \left( (\tau_1 + \tau_2)z_n b_n + \tau_3\right) b_n + \tau_0 \right) e_n, \lab{W_e_AW} \ea
where $b_n, u_n$ coincide with the recurrence coefficients of the big $q$-Jacobi polynomials \re{ub_J}.

Using an appropriate similarity transformation
\be
\tilde W = S W S^{-1}, \lab{t_W} \ee
with $S$ diagonal:
\be
S e_n = s_n e_n, \lab{S_s} \ee
we can bring the operator $\tilde W$ in the ``monic'' form
\be
\tilde W = e_{n-1} + \tilde b_n e_n + \tilde u_{n+1} e_{n+1}, \lab{tW_monic} \ee
where 
\ba
&&\tilde b_n = \tau_0 + \left( (\tau_1 + \tau_2)z_n b_n + \tau_3\right) b_n, \nonumber \\
&&\tilde u_n = (\tau_1 z_{n-1} + \tau_2 z_n + \tau_3)(\tau_1 z_{n} + \tau_2 z_{n-1} + \tau_3) u_n. \lab{t_bu_W} \ea
If one chooses 
\ba
&&c_1=a_1a_2q^{-1}, \: c_2=a_3a_4q^{-1}, \: c_3=a_1a_3q^{-1}, \: \tau_1 = \frac{1}{2a_1a_2a_3 (q-q^{-1})}, \nonumber \\
&&\tau_2 =-q \tau_1, \: \tau_3 = \frac{1}{2 a_1}, \: \tau_0= \frac{q(a_2+a_3) + a_2a_3(a_1+a_4)}{2(q+1)a_2a_3}, \lab{tau_AW} \ea
it follows that
\be
\tilde u_n = u_n^{(AW)}, \quad \tilde b_n = b_n^{(AW)} \lab{tub_AW} \ee
and $W$ hence coincides (up to the similarity transformation \re{t_W}) with the operator giving the three-term recurrence relation of the Askey-Wilson polynomials.

This is how these polynomials can be obtained by tridiagonalization from the  big $q$-Jacobi polynomials. 
Note that the Wilson polynomials were derived similarly from the ordinary Jacobi polynomials  \cite{GIVZ}.

Let us return to the $q$-oscillator algebra representation with $A$ corresponding to the Jacobi matrix associated to the big $q$-Jacobi polynomials and $B$ as in \eqref{B_A_J}. 
Consider the operator $C$ given by
\be
C = A + \mu B + \lambda \mathbb{I} \lab{C_pencil} \ee
where $\lambda, \mu$ are arbitrary real parameters.
It is clear that $C$ is also tridiagonal and it is readily seen to act as follows
\be
C e_n = V_{n+1}^{(1)} e_{n-1} + V_n^{(2)} e_n + V_n^{(3)} e_{n+1}, \lab{C_3_J} \ee
where
\ba
&& V_n^{(1)}= 1+ \mu r_2 z_{n-1} - q r_2 z_n, \; V_n^{(2)}= r_0 \mu + \lambda + b_n (1+\mu (1-q)r_2 z_n), \nonumber \\
&&V_n^{(3)}= (1+ \mu r_2 z_{n+1} - q r_2 z_n) u_{n+1}. \lab{V_J} \ea
It is manifest in view of  \eqref{B_A_J} that the linear pencil \re{C_pencil} is in the class of the operators $W$ that were obtained above by the tridiagonalization \re{W_A}.
It thus follows that linear combinations of the two generators $A$ and $B$ of the $q$-oscillator algebra in the tridiagonal representation have the Askey-Wilson polynomials as eigenfunctions for generic choices of the parameters $\mu$ and $\lambda$.

\section{Finite-dimensional reductions: \\ $q$-Hahn and $q$-para Krawtchouk polynomials}
\setcounter{equation}{0}
So far, the  operators $A$ and $B$ were assumed to act on infinite-dimensional spaces. 
This led to a picture where the big $q$-Jacobi polynomials showed up as eigenvectors of the operators $A$ and $B$ and where the Askey-Wilson polynomials arose as eigenvectors of the linear pencil $A+ \mu B + \lambda \mathbb{I}$ with arbitrary parameters $\lambda$ and $\mu$.

We shall now consider two finite-dimensional reductions. One will have the $q$-Hahn polynomials as eigenvectors of the Jacobi matrix $A$ and the $q$-Racah polynomials as eigenvectors of the pencil $A+ \mu B + \lambda \mathbb{I}$. The truncation condition is then the standard one:
\be
c_3=q^{-N-1}, \quad N=1,2,\ldots \lab{tr_Hahn} \ee
which gives the $q$-Hahn polynomials as the finite-dimensional companions of the big $q$-Jacobi polynomials \cite{KLS}.
The spectrum of the operator $A$ is in this case exponential: 
\be
x_s = q^{-s}, \; s=0,1,\dots, N, \lab{x_s_Hahn} \ee
in keeping with the results on finite-dimensional irreducible representations of the $q$-oscillator algebra \cite{MS} according to which  the spectrum of any of the operators $A$ or $B$ is a simple geometric sequence.

Another possible finite-dimensional reduction leads to orthogonal polynomials that to our knowledge were never considered so far, namely,  
the $q$-para Krawtchouk polynomials as eigenvectors of the matrix $A$ and the $q$-para Racah polynomials as eigenvectors of operator pencil $A+ \mu B + \lambda \mathbb{I}$.
These are obtained when $N$ is odd under the truncation condition
\be c_1 = c_2 = q^{-(N+1)/2}. \lab{tr_para} \ee
The polynomials $P_n(x)$ that are obtained from the big $q$-Jacobi ones by imposing \eqref{tr_para} satisfy the three-term recurrence relation
\be
P_{n+1}(x) + (1-D_n - C_n) P_n(x) + D_{n-1} C_n P_{n-1}(x) = x P_n(x), \lab{rec_para} \ee
where
\be
D_n = {\frac { \left( 1-{q}^{n-N} \right)  \left( 1-c_3\,{q}^{n+1}
 \right) }{ \left( 1-{q}^{2\,n-N} \right)  \left( 1+{q}^{n-(N-1)/2}
 \right) }} \lab{A_para} \ee
and
\be
C_n = -c_3\,{q}^{n-(N-1)/2} \frac{\left( 1-{q}^{n} \right)  \left( 1-c_3^{-1}
{q}^{n-N-1} \right) }{ \left( 1-{q}^{2\,n-N} \right) 
 \left( 1+{q}^{n-(N+1)/2} \right)}. \lab{C_para} \ee
These polynomials have an unusual spectrum which is the superposition of two exponential sublattices:
\be 
x_{2s} = q^{-s}, \quad x_{2s+1}= c_3 q^{s+1}, \quad s=0,1,\dots, \frac{N-1}{2}.  \lab{para_spec} \ee 
These spectral points are the zeros of the characteristic polynomial $P_{N+1}(x)$:
\be
P_{N+1}(x_s) =0, \quad s=0,1,\dots, N. \lab{root_N+1} \ee
When $q \to 1$, the polynomials $P_n(x)$ tend to the para Krawtchouk polynomials introduced in \cite{VZ}, which have a spectrum consisting of two uniform sublattices. 
It is therefore natural to coin the name $q$-para Krawtchouk for the polynomials $P_n(x)$.

These two examples of finite-dimensional reduction of the tridiagonal representation of the $q$-oscillator algebra  can be put in perspective with the general results on the irreducible finite-dimensional representations of this algebra. 
(An irreducible representation is one that does not have an invariant subspace.)  
It was shown in \cite{MS} that for any irreducible finite-dimensional representation of the $q$-oscillator algebra, 
the matrices $A$ and $B$ can be cast in the canonical form $A_c, B_c$ where the matrix $A_c$ is diagonal
$$
A_c=\left[
\begin{array}{ccccc}
a & 0 & 0 & \cdots & 0 \\
0 & a q^{-1} & 0 &  & \vdots  \\
0 & \ddots & \ddots & \ddots & 0 \\
\vdots & & 0 & a q^{1-N} & 0 \\
0 & \cdots & 0 & 0 & a q^{-N}
\end{array}
\right]$$
and $B_c$ is bidiagonal 
$$
B_c=\left[
\begin{array}{ccccc}
a' & 1 & 0 & \cdots & 0 \\
0 & a' q & 1 &  & \vdots  \\
0 & \ddots & \ddots & \ddots & 0 \\
\vdots & & 0 & a' q^{N-1} & 1 \\
0 & \cdots & 0 & 0 & a' q^{N}
\end{array}
\right] ,
$$
with $a$ an arbitrary nonzero real parameter and $a a' (1-q)=1$. 

The Jacobi matrices $A$ and $B$ that correspond to the $q$-Hahn polynomials are irreducible and there thus exists a similarity transformation $S$ that transforms the matrices $A$ and $B$ into $A_c$ and $B_c$:
\be
S A S^{-1} = A_c, \quad S B S^{-1} = B_c. \lab{sim_0} \ee
The spectrum of the matrix $A$ is in this case  $x_s=q^{-s}, s=0,\dots, N$ implying that the parameter $a$ in $A_c$ is equal to 1. 
Note that $a$ is not essential as it can be removed by the trivial automorphism $A \to a A, \: B \to a^{-1} B$ of the $q$-oscillator algebra.

For the $q$-para Krawtchouk polynomials, the spectrum is a combination of two geometric series. This means that the corresponding representation of the $q$-oscillator algebra is completely reducible: it can indeed be brought under a similarity transformation into the direct sum of two tridiagonal matrices each having, as $A_c$, single geometric series as spectrum.

\section{Algebraic relations}
\setcounter{equation}{0}
Consider the 3 operators $A, B$ and $Z$ introduced in Section 3. 
It is easily verified that they satisfy the following relations
\ba
&&AB - q BA = \mathbb{I}, \quad BZ - q Z B = \gamma_1 A + \delta_1 \mathbb{I}, \quad ZA - q AZ = \gamma_2 B + \delta_2 \mathbb{I}, \lab{big_AW} \ea
where
\ba
&&\gamma_1 = -\frac{c_2(q+1)}{c_3 q}, \; \delta_1 = c_2 c_3^{-1}(c_1+1) + c_2+1, \nonumber \\
&&\gamma_2= -c_1c_3 q (q+1) (1-q)^2, \; \delta_2 = -(q+1) \left(c_1(c_2+1) +c_3(c_1+1) \right) . \lab{gg_big} \ea
Note that the first relation in \re{big_AW} is the defining relation of the $q$-oscillator algebra while the third relation is equivalent to \re{B_A_J}.

The algebraic relations \re{big_AW} can be considered as a special case of the Askey-Wilson algebra introduced in \cite{ZheAW} (see also \cite{WZ}, \cite{TUAW}):
\ba
&&AB - q BA = \gamma_0 Z + \delta_0 \mathbb{I}, \quad BZ - q Z B = \gamma_1 A + \delta_1 \mathbb{I}, 
\quad ZA - q AZ = \gamma_2 B + \delta_2 \mathbb{I} \lab{AWR} \ea
with arbitrary parameters $\gamma_i, \delta_i, \: i=0,1,2$. 
If  $\gamma_0 \gamma_1 \gamma_2 \ne 0$ the algebra \re{AWR} can be brought  by appropriate scaling transformations of the operators $A,B,Z$, to a ``canonical'' form \cite{TUAW} with $\gamma_0=\gamma_1=\gamma_2=1$. 
The algebra \re{AWR} gives an algebraic description of the Askey-Wilson polynomials \cite{ZheAW}.

The Askey-Wilson algebra is degenerate in the present case because $\gamma_0=0$, it here encodes the properties of the big $q$-Jacobi polynomials and will be called the big $q$-Jacobi algebra. 
(An equivalent form of the AW-algebra corresponding to the big $q$-Jacobi polynomials was considered in \cite{Ma}.)
Note that if one puts $q=-1$ in \re{big_AW}, we obtain the anticommutator algebra
\ba
&&AB + BA = \mathbb{I}, \quad BZ + Z B = \gamma_1 A + \delta_1 \mathbb{I}, \quad ZA + AZ = \gamma_2 B + \delta_2 \mathbb{I} \lab{big-J} \ea
which was considered in \cite{VZ2}.
This last algebra corresponds to the big -1 Jacobi polynomials.

Let us now revisit the algebras that are formed when instead of starting with $A$, one begins with 
\be L=A + \mu B \lab{LAB} \ee
which coincides with the operator pencil \re{C_pencil}.
 (No generality is lost by setting $\lambda=0$  since this parameter simply shifts all eigenvalues.)
To $L$ and $Z$, add the operator $M$
defined by
\be M=LZ - q ZL - \omega_0 \mathbb{I}, \lab{M_def} \ee
where
\be \omega_0= q(q-1) \left( c_3(c_1+1) + c_1 (c_2+1) \right) - \mu c_3^{-1}  \left( c_3(c_2+1) + c_2 (c_1+1) \right). \lab{omega0} \ee
We  then have
\ba
ZM-qMZ = \sigma_1 L + \omega_1  \mathbb{I}, \quad MK-qLM = \sigma_2 Z + \omega_2  \mathbb{I} \lab{AW_rels} \ea
where
\ba
&&\sigma_1= c_1c_2 (q^2-1)^2, \quad \omega_1= c_2 c_3^{-1}\mu (q^2-1) \left( c_3(c_1+1) + c_1 (c_2+1) \right)- \nonumber \\
&&  -c_1 q (q+1)(q-1)^2   \left( c_3(c_2+1) + c_2 (c_1+1) \right)  \lab{so1} \ea
and 
\ba
&&\sigma_2= \frac{\mu(1-q)(q+1)^2}{q}, \quad \omega_2= \mu (q^2-1)  \left( c_1 c_2(c_3^{-1}+1) + c_1 +c_2+c_3+1 \right) - \nonumber \\
&&  c_1c_3 q(q+1)(q-1)^2 - \frac{\mu^2 (q+1)c_2}{qc_3}. \lab{so2} \ea
It is then seen that the operators $L,M$ and $Z$ satisfy commutation relations of the type \re{AWR}. 
A realization of the generic Askey-Wilson algebra \re{AWR} is thus obtained from the tridiagonal $q$-oscillator algebra representation when the linear pencil $L=A +\mu B$ is taken as one of the generators. 
Tridiagonalization was used in \cite{GIVZ} to arrive at the Wilson polynomials from the Jacobi ones and it was shown in that reference that the Racah-Wilson algebra can be built from the Jacobi algebra. 
The construction of the Askey-Wilson algebra from the linear pencil $A+\mu B$ of tridiagonal $q$-oscillators provides a $q$-generalization of the results obtained in \cite{GIVZ}.

\section{Concluding remarks}
Summing up, we have found the most general tridiagonal representations of the $q$-oscillator algebra under the assumption that the generators are irreducible.
The operators $A$ and $B$ then correspond to Jacobi matrices of the big $q$-Jacobi polynomials and the connection with the Askey-Wilson polynomials has found a natural explanation through tridiagonalization. A conclusion is that the 
tridiagonal representations introduced in the context of ASEP model studies \cite{USW} are generic and that the Askey-Wilson polynomials are the most general ones that can be associated to an element of the from $A + \mu B$.

Our study raises interesting questions.
It should prove worthwhile to study in detail the generalized eigenvalue problem $A + \mu B + \lambda \mathbb{I} =0$ for $\lambda$ fixed and to determine the biorthogonal rational functions that are the eigenvectors \cite{DVZ}.
Looking for an explicit automorphism that would transform the standard representation with upper and lower subdiagonal matrices into the tridiagonal one is certainly of relevance. 
This should bear a relation with the bootstrapping approach to Askey-Wilson polynomials given in \cite{Atakishiyev2010,Atakishiyev2011}. 
Finally, what is the relation, if any, between the algebraic description of the Askey-Wilson polynomials based on the $q$-oscillator algebra and the one resting on the rank one DAHA of type $(C_{1}^{\vee}, C_1)$?
We hope to report on these questions in the future.

\bigskip\bigskip
\noindent
{\Large\bf Acknowledgments}
\bigskip
\\
LV wishes to thank S.~Sahi for stimulating discussions.
AZ acknowledges the hospitality of Kyoto University during the course of this work.
The research of LV is supported in part by a grant from the Natural Sciences and Engineering Research Council (NSERC) of Canada and that of ST by JSPS KAKENHI Grant Numbers 25400110.

\bibliographystyle{amsplain}

\end{document}